\documentclass[12pt]{article}

\newcommand{\be}{\begin{equation}}
\newcommand{\ee}{\end{equation}}
\newcommand{\bea}{\begin{eqnarray}}
\newcommand{\eea}{\end{eqnarray}}
\newcommand{\nn}{\nonumber}

\begin{document}

\title{Effects of Magnetic Fields on String Pair Creation}

\author{Ciprian Acatrinei\thanks{e-mail acatrine@sissa.it;
        on leave from \it{Institute of Atomic Physics -
        P.O. Box MG-6, 76900 Bucharest, Romania}}\\
        {\it SISSA - Via Beirut 2-4, 34014 - Trieste, Italy} }

\date{January 2000}

\maketitle

\begin{abstract}
The rate of pair production of open strings in general uniform electromagnetic fields
is calculated in various space-time dimensions.
The corrections with respect to the case of pure electric backgrounds are displayed.
In particular, a contribution in the form of a Born-Infeld action is derived and its role 
in the present context emphasized
\end{abstract}

%\noindent
\section{Introduction and summary}

One way to unravel some nonperturbative features of quantum systems 
is to study their dynamics in the presence of background fields.
A classic example is Schwinger's calculation \cite{schwinger}
of the rate at which electron-positron pairs are created by a
%tunneled out of the vacuum in the presence of a
constant electric field. 

More recently, the dynamics of linearly extended objects placed
in various backgrounds became of interest in the context of string theory. 
In this direction,
%the classical and quantum dynamics of 
open strings coupled
through their end-points to external electromagnetic fields
have
%has 
been studied by various authors \cite{ft,acny,burgess,nesterenko,bp}. 
In particular, the Born-Infeld \cite{bi} effective action 
for the external field was obtained in \cite{ft,acny},  
whereas \cite{burgess,bp} discussed the possibility 
of string pair creation through the Schwinger mechanism.
Ref. \cite{burgess} considered bosonic strings in weak electric fields, 
whereas Ref. \cite{bp} studied both bosonic and fermionic strings, 
for electric field strengths up to the value (in natural units) of the string tension.  
The above quantum effects were studied by calculating a one-loop effective action 
(whose imaginary part is relevant for the pair creation process) 
in a background electric field;
additional magnetic fields were left out of this type of analysis,
although they affect the pair creation rate. 

The present note is an extension and a generalization of the seminal work \cite{bp}.
We study the creation of open strings in presence of a constant, 
but otherwise arbitrary, electromagnetic field strength $F_{\mu\nu}$,  
and compute the pair production rate (PPR) in this general situation. 
We display the various corrections with respect to 
the pure electric field background result, 
which is recovered as a particular case. 

Moreover, our results will be obtained via path integral methods.
Their use offers, even in the case of pure electric fields,
several advantages over the Hamiltonian methods used in \cite{bp}:
the derivation is self-contained; it does not rely on previous work \cite{acny}
concerning the spectrum of the open string in external fields, but rather encodes
it in the final result. In this way one also bypasses a somehow puzzling aspect of canonical
methods, namely the use of a complex valued substraction for the vacuum, and hence for
the spectrum, of the string in electric fields. 
For arbitrary external electromagnetic fields, which are our concern,
path integrals allow a direct calculation - without previous diagonalization of $F_{\mu\nu}$ -
of the PPR. This is difficult to envisage through canonical methods.
Still, to compare easily with the work of \cite{bp}, we will choose to first 
block-diagonalize $F_{\mu\nu}$. This will ask for a semiquantitative analysis of the 
dependence of the eigenvalues of $F_{\mu\nu}$ on its non-diagonal components,
which is of general interest for physics in more than four dimensions.
It will turn out that one cannot find the eigenvalues analytically for a spacetime 
dimension $D \geq 10$.

We will also provide a transparent derivation of the Born-Infeld (BI) effective action, 
complementary to the original work of \cite{ft,acny}. 
In particular,
the connection between the BI action and the finite spatial extension of 
the string is clearly seen in this formalism, 
as already noticed in \cite{ij2} (Ref. \cite{ai} anticipated
our present discussion in the case of pure electric fields). 
Beyond being somehow at the center of the calculation,
the BI term produces an enhancement of the PPR,
a qualitative stringy effect somehow unexpected in a bosonic theory
(where the magnetic field usually decreases the pair production rate).
%[but try to see spin one first!]
This effect becomes particularly important in the dissipative limit
of string theory described in \cite{ij2, ai}.
We treat in detail only the case of bosonic strings, 
which displays all the interesting aspects of the situation;
the results for the case of superstrings are only briefly mentioned.
Our methods and results might be of some use in the
light of D-branes \cite{polbach}, for instance in the context of
non-commutative low energy actions for strings 
stretched between branes with electromagnetic fluxes on them \cite{ncg}.  
The results obtained in this note were checked by using 
the boundary state formalism \cite{bsf}.

%instabilitati: nu mai bine aici scurt si pe larg cand le discuti?

Last but not least,
the open string spectrum in presence of a magnetic field
contains tachyonic excitations \cite{fp}, which destabilize the theory.
These instabilities, 
like the similar tachyonic mode appearing in Yang-Mills theories \cite{no},
can be traced back to the
nonminimal coupling of the electromagnetic field to 
excitations with spin greater than one \cite{gs_2}.
Our expedient remedy will be to 'Higgs' the theory 
\footnote{I am grateful to C.Bachas for useful discussions on this point} 
by stretching the string between some different, parallel, D-branes. 
The Dirichlet boundary conditions
along the coordinates orthogonal to the branes
then produce an additional mass term, 
%(proportional to the string tension multiplied by the square of the interbrane distance).
which can overcome the destabilizing 
contribution of the magnetic field.
%; it also affects the PPR. 

%The instability produced by the magnetic field is similar  
%which questioned seriously 
%the validity of the so-called Savvidy vacuum, characterized by a 
%minimal value of the 1-loop effective potential at non-zero values 
%of the external (chromo)magnetic field.

%\bigskip
%\noindent
\section{Generalities}

{\it i) Vacuum decay rate.} We briefly review one way to calculate the rate 
at which pairs are tunneled out of the vacuum, in presence of an external field. 
%Although the method is quite general, we will restrict our attention
%to the case in which the relevant degreees of freedom are open strings.

The vacuum free energy in 
%presence of 
an external 
%electromagnetic 
field $F_{\mu\nu}$ is the logarithm of
the vacuum-to-vacuum transition amplitude
$ e^{-i W(F_{\mu\nu})_{(vac)} }=<0|e^{-i\hat{H}\times (time) }|0>$.
For a static field, $W_{vac}(F_{\mu\nu})={\cal E}_{vac}(F_{\mu\nu}) 
\times (time)$, where $(time)$ is the total time interval.
Reexpressing the vacuum free energy {\it \`{a} la} Schwinger, 
and taking the Hamiltonian to be the one for open strings, 
one obtains
\begin{equation}
W_{vac}=\int_{0}^{\infty}\frac{dt}{t} 
Tr e^{-t \hat{H}_{string}}.
\end{equation}
The trace is evaluated by means of a (suitably normalised) path integral
\be
Tr e^{-t \hat{H}_{string}}=\int DX_{0}D\vec{X} e^{-S(X_{0}, \vec{X}, F_{\mu\nu})}. \label{pi}
\ee
The action $S(X_{0}, \vec{X}, F_{\mu\nu})$ 
(discussed below) 
includes the external electromagnetic field, 
which couples to the end-points of the string.
Due to that coupling, the vacuum energy
${\cal E}_{vac}(F_{\mu\nu})$
%$=Re {\cal E}_{vac}(F_{\mu\nu})-i\frac{\Gamma}{2}$, 
gets an imaginary part, $\frac{\Gamma}{2}$,
which induces the vacuum decay.
The decay rate per unit volume $\gamma =\frac{\Gamma}{V}$
%is obtained by discarding the
%zero modes of the coordinates on which we path-integrate in (\ref{pi})
%(they give precisely the space-time volume) and 
reads
\begin{equation}
\gamma=-2 Im\int_{0}^{\infty}\frac{dt}{t}\int' 
DX_{0}D\vec{X}e^{-S(X_{0},\vec{X},F_{\mu\nu})}. \label{rate}
\end{equation}
The prime means that we have factored out the zero mode part of the action
(upon integration, it gives precisely the space-time volume).
%, as will be tacitely understood in what follows. 

For a bosonic open string living on a Euclidean world-sheet, 
coupled to a $U(1)$ gauge field, the action to be used in eqs.(\ref{pi},\ref{rate}) reads:
 
\bea 
S & = & \frac{T}{2}\int_{0}^{t} d\tau \int^{l}_{0}
        d\sigma[(\frac{\partial X^{\mu}}{\partial \tau})^{2}
        +(\frac{\partial X^{\mu}}{\partial \sigma})^{2}]  \label{action1} \\
  &   & -i q_{1} F_{\mu\nu}\int_{0}^{t} d\tau \left[X_{\mu}
        \frac{\partial X_{\nu}}{\partial \tau}\right]_{\sigma=0}     
        -i q_{2} F_{\mu\nu}\int_{0}^{t} d\tau \left[X_{\mu}
        \frac{\partial X_{\nu}}{\partial \tau}\right]_{\sigma=l}, \nn 
\eea 
where $T$ denotes the string tension, whereas $q_{1}$ and $q_{2}$ are the magnitudes 
of the charges situated at the end-points of the string.
This action being invariant under a rescaling of both $\sigma$ and $\tau$ by a factor of $l$, we set $l=1$.
%We can set $l=1$.

\bigskip
\noindent
{\it ii) Free path integral.}
%Once we have the action $S$, we can start 
%evaluating the path-integral in eq.\ref{pi}.
%Before analyzing the path-integral and vacuum amplitude in various dimensions,
We first review the zero electromagnetic field case \cite{ai}.
If $F_{\mu\nu}\equiv 0$, eq.(\ref{pi}) factorizes into products of  
free path integrals along each space-time direction.
For a generic uncoupled coordinate $X$ we have to evaluate
$\int DX e^{-S}$, with
$ S=\frac{T}{2} \int_{0}^{t} d\tau \int_{0}^{1} d\sigma 
 [(\frac {\partial X}{\partial \tau})^{2}+(\frac{\partial X}{\partial \sigma})^{2}]$.
Taking the  boundary conditions 
to be periodic along $\tau$, $ X(t+\tau,\sigma)=X(\tau,\sigma)$, 
and Neumann along $\sigma$ ,   $\frac{\partial X}{\partial \sigma}|_{\sigma=0,1} =0,$
which means expanding $X(\tau, \sigma)$ as follows
\begin{equation}
X(\tau,\sigma)=\sum_{n\in Z}\sum_{k\in N}X_{nk}cos(k\pi\sigma)
exp(2\pi in\frac{\tau}{t}), \label{expansion1}
\end{equation}
one gets \cite{ai} the result
\be
Z_{free}\equiv \int DX e^{-S}=\sqrt{\frac{T}{4\pi}}
%\left ( e^{-\frac{\pi}{2}\frac{t}{l}\sum_{k>0}k} 
%e^{-\frac{\pi}{12}\frac{t}{2l}} \right )
e^{\frac{\pi}{6}\frac{1}{t}}
\prod_{n \geq 1}\frac{1}{1-e^{-4\pi n \frac{1}{t}}}. \label{free1}
\ee
Through a modular transformation 
one can reexpress (\ref{free1}) as follows: 
\be
Z_{free}=\sqrt{\frac{T}{4\pi}\frac{2}{t}}
e^{\frac{\pi}{24}t}
\prod_{k\geq 1}\frac{1}{1-e^{-\pi k t}}. \label{free2}
\ee

\bigskip
\noindent
{\it iii) 'Higgsing' the theory.}
Alternatively, one could use Dirichlet boundary conditions for the string's ends,
by obliging them to stay on two parallel D-branes situated at 
a relative distance $d$, i.e. ask $X(\sigma =0)=0$ and $X(\sigma =1)=d$. 
The corresponding mode expansion is

\begin{equation}
X(\tau,\sigma)=d\sigma+\sum_{n\in Z}\sum_{k>0}X_{nk}\sin(k\pi\sigma)
exp(2\pi in\frac{\tau}{t}) ,
\end{equation}
and it 
modifies the final result of the path integration in only one way: 
an additional
exponential factor $e^{-\frac{t}{2}Td^2}$ appears in (\ref{free1}). 
This is equivalent to a mass term $\frac{1}{2}Td^2$,
whose role is to stabilize the theory in presence of a magnetic field. 
This term can be inserted at any step of the calculation; 
we will come back to it later.

%\bigskip
%\noindent
\section{Eigenvalues}
%The critical dimension for a bosonic string is $26$, but the electromagnetic
%field strengths can be restricted to a subspace, of dimension four, 
%for instance.
%On the other end, one can compactify some of the free directions..
%(interesting to see what happens if compactify some direction with
%an electric field along, or $B_{ij}$, $X_{i}$ non-compact and $X_{j}$
%compact)
%In fact, we will not care about the critical dimension.
A non-zero background field $F_{\mu\nu}$ can be block-diagonalized
in any dimension $D$:
$F_{01}=-F_{10}=\cal E$, $F_{23}=-F_{32}={\cal B}_{1}$, $F_{45}=-F_{54}={\cal B}_2$, etc.,
and $F_{ij}=0$ for $i\neq j \underline{+}1$.
%\be
%F=\left (
%\begin{array}{ccccccc}
%0 & \cal E & 0 & 0 & 0 & 0 & \cdot  \\
%-\cal E & 0 & 0 & 0  & 0 & 0 & \cdot \\
%0 & 0 & 0 & {\cal B}_{1}  & 0 & 0 & \cdot \\
%0 & 0 & -{\cal B}_{1} & 0 & 0 & 0 & \cdot \\
%0 & 0 & 0 & 0 & 0 & {\cal B}_{2} & \cdot \\
%0 & 0 & 0 & 0 & -{\cal B}_{2} & 0 & \cdot \\
%\cdot & \cdot & \cdot & \cdot & \cdot & \cdot & \cdot    
%\end{array}
%\right ),
%\ee
${\cal E}$ represents the electric-like eigenvalue of $F_{\mu\nu}$,
%(equal to the electric field when all the initial $F_{1j}$'s, with $j\neq 0,1,$ do vanish)
whereas the ${\cal B}$'s are the magnetic-like ones.
%In odd dimensions the last eigenvalue is zero.
The PPR rate being a relativistic invariant, one can calculate it
either by first diagonalizing $F_{\mu\nu}$ and path integrating subsequently, 
or by path integrating directly.
The second approach will be mentioned later (eq. \ref{nodiag}).
We will use a block-diagonalized $F_{\mu\nu}$ 
%matrix
and 
we will 
obtain the PPR rate as a function of its eigenvalues. 
It is thus of interest to study their dependence 
on the initial, in general non-diagonal, field strength.

%The complete filling of the $F_{\mu\nu}$ matrix has two effects:
%first, it produces non-zero magnetic eigenvalues, which have a
%moderate impact on the PPR; they enter into
%a prefactor, in the form of infinite products. 
The most important effect of non-diagonal $F_{ij}$'s 
is to change $\cal E$ - which enters the exponential factor of the PPR.
Inserting an imaginary factor in front of the electric components $F_{0j}$,
in order to use the Euclidean metric in the eigenvalue equation 
$det(F_{\mu\nu}-\eta_{\mu\nu}\lambda)=0$,
$F_{\mu\nu}$ reads, in $D=4$,
 
\be
F=\left(
\begin{array}{cccc}
0 & i E & 0 & 0 \\
-i E & 0 & b & 0 \\
0 & -b & 0 & B \\
0 & 0 & -B & 0
\end{array}
  \right). \label{f4}
\ee
$B$ and $b$ are the components of the magnetic field parallel, 
respectively orthogonal, to the electric field $E$; 
$\cal E$, the real eigenvalue of $F$, is
%Then, the characteristic equation is
%\be
%Det[\lambda I-F]=\lambda^4+(B^2+b^2-E^2)\lambda^2-E^2 B^2=0
%\ee    
%with roots
\be
2({\cal E}^2)_{1,2}=\sqrt{(B^2+b^2-E^2)^2+4E^2B^2}-(B^2+b^2-E^2).
\ee
For $b=E$, 
%to take the most sugestive case, 
%the electric eigenvalue (the real one) 
${\cal E}$ decreases with respect to the case $b=0$; 
%(i.e. parallel electric and magnetic fields), 
it vanishes if $B=0$.
%If one more dimension is allowed, more dramatic effects enter into play.
In five dimensions, for a field strength tensor with non-zero components
%\be
%F=\left(
%\begin{array}{ccccc}
%0    & i E   & 0  &   0 & 0 \\
%-i E & 0     & b  &   0 & 0 \\
%0    & -b    & 0  &   0 & 0 \\
%0    & 0     & 0  &   0 & B \\
%0    & 0     & 0  & - B & 0
%\end{array}
%  \right),  \label{f5}
%\ee 
\be
F_{01}=iE, \quad F_{12}=b, \quad F_{34}=B, \label{f5}
\ee
and if $b=E$, the electric-like eigenvalue is zero for any $B$. 
This happens because, inside the matrix given by (\ref{f5}),
the block containing $E$ and $b$ does not have lines or columns 
in common with the block containing the $B$'s.
%(If two such $b$'s, say $b_1$ and $b_2$, are allowed, e.g. in $D=6$, 
%${\cal E}$ vanishes even quicker, at $b^2_1+b_2^2=E^2$.)

In general, for an electric field parallel to the $x$-axis
and a purely magnetic block partially diagonalized (with the $F_{1j}$'s
left in their original form)
%, i.e. for a matrix of the form

\be
F=\left (
          \begin{array}{ccccccc}
          0 & iE & 0 & 0 & 0 & 0 & \cdot\\
          -iE & 0 & b_1 & b_2 & b_3 & b_4 & \cdot\\
          0 & -b_1 & 0 & B_1 & 0 & 0 & \cdot\\
          0 & -b_2 & -B_1 & 0 & 0 & 0 & \cdot\\
          0 & -b_3 & 0 & 0 & 0 & B_2 & \cdot\\
          0 & -b_4 & 0 & 0 & -B_2 & 0 & \cdot \\
          . & .   & . & . & .   & . & . 
          \end{array} \label{f6}
  \right ),
\ee
%(the other $b$'s being superfluous in $6$ dims),
%: $E=F_{01}\neq 0$,
%$F_{0i}=0; \forall i\neq 1$, 
the magnetic components fall into two classes: 
1) The $F_{1j}$'s, called here $b$'s; 
they decrease the electric eigenvalue $\cal E$ and 
might even cancel it (as in eq.\ref{f5}),
unless they sit above a $2\times 2$ block
containing a non-zero $B$ (as in eq.\ref{f4}). 
2) The other components, the $B$'s; 
%they have a less important role and
they do not influence ${\cal E}$ in absence of the $b$'s;
for non-zero $b$'s, they temper the reduction of $\cal E$ those produce. 
At fixed $F_{1j}$'s, $\cal E$ grows when the $B$'s are increased. 
This is easily seen up to $D=5$ space-time dimensions and, 
in principle, also  up to $D=9$, by finding analytic expressions for the eigenvalues. 
%(although this is of little practical value for $D\geq 6$).
In ten dimensions the characteristic equation of the matrix $F$ 
becomes of degree five and is not any more solvable by radicals. 
We have tested numerically various cases for $D$ from $6$ to $10$, 
and the conclusions above held. 
The $F_{1j}$'s reduce $\cal E$ and the production rate, whereas the other 
magnetic components temper their decreasing effect if their corresponding 
planes intersect. This is probably true in any space-time dimension.
Moreover, in higher dimensions the effect of a given variation 
of one single $F_{ij}$ is less important 
than a similar change in lower dimensions.
The other numerous components provide a kind of inertia. 
%one should change many $B$'s to get a sensible effect.

One can also fill the empty off-diagonal magnetic part of $F$
(e.g. the $4\times 4$ matrix containing $B_1$ and $B_2$ in (\ref{f6})).
Increasing those components might decrease or increase $\cal E$,
but their influence is small, being supressed by at least one order of magnitude 
with respect to their initial variation.

%We end here the discussion of general matrices, and 
%- keeping in mind the effect of non-diagonal terms on ${\cal E}$ -
%proceed to the explicit calculation of the PPR 
%for $F_{\mu\nu}$ already put into block-diagonal form.

%\bigskip
%\noindent
\section{Path integral evaluation}

Once $F_{\mu\nu}$ is diagonalized, 
we are left with path integrals along pairs of coupled coordinates. 
At this point, we could just use the results obtained in \cite{bp}, for pure electric fields.
Nevertheless, we prefer to make a more direct, path integral, calculation. 
Beyond being of intrinsic interest, the evaluation of the path integral which follows 
is a different way to make one of the few known nonperturbative, nonsupersymmetric, 
but still interesting, calculations in string theory.
Choosing the $0-1$ plane, we evaluate

\be
Z_{01}=\int DX_{0}DX_{1} e^{-S(X_{0}, X_{1}, F_{01})},
\ee
the action $S(X_{0}, X_{1}, F_{01})$ being eq. (\ref{action1}) now
restricted to $\mu =0,1$:

\bea 
S & = & \frac{T}{2}\int_{0}^{t} d\tau \int^{1}_{0}
        d\sigma[(\frac{\partial X^{\mu}}{\partial \tau})^{2}
        +(\frac{\partial X^{\mu}}{\partial \sigma})^{2}]  \label{action2} \\  
  &   & -i q_{1} E_{1}\int_{0}^{t} d\tau \left[X_{0}
        \frac{\partial X_{1}}{\partial \tau}\right]_{\sigma=0}    
        -i q_{2} E_{2}\int_{0}^{t} d\tau \left[X_{0}
        \frac{\partial X_{1}}{\partial \tau}\right]_{\sigma=1}. \nn 
\eea
We remark that we can treat in this way 
the more general case in which different field strengths $E_{1,2}$ are applied 
to the two string end-points. 
We will subsequently include the two charges $q_1$ and $q_2$ into the field 
strength value through the more compact notation $q_{1,2}E_{1,2}\rightarrow E_{1,2}$.

Developing in the same interaction-independent 
Fourier basis as in the free case (\ref{expansion1}),
%\be
%X(\tau,\sigma)=\sum_{n\in Z}\sum_{k\in N}X_{nk}
%cos(k\pi\frac{\sigma}{l})exp(2\pi in\frac{\tau}{t})
%\ee
the action (\ref{action2}) becomes
$ S=S(0)+\sum_{n=1}^{\infty}S(n) $,
with 
$ S(0)=-\frac{t}{2} \sum_{k>0}[(X^{0}_{0k})^{2} 
\frac{T}{2}\pi^{2} k^{2}-
(X^{1}_{0k})^{2} \frac{T }{2}\pi^{2} k^{2}] $
and
$ S(n>0)= {\bf X^{\dagger} }A{\bf X} $.
The term ${\bf X^{\dagger}}A{\bf X}$ 
encodes the modes which couple due to the electric field: 
%Through the notation
$$ 
{\bf X^{\dagger}}=( 
X^{0}_{-n,0}, X^{0}_{-n,1}, \dots X^{0}_{-n,k}, \dots , 
X^{1}_{-n,0}, X^{1}_{-n,1}, \dots X^{1}_{-n,k}, \dots ),
$$

%the matrix $A$ is given by

\be
%gin{displaymath}
A = \left ( 
\begin{array}{ccccccccccccc}
a_{0} & 0 & 0 & \cdot & \cdot &  &  C_{1} & C_{2} & C_{1} & \cdot & \cdot &  \\
0 & a_{1} & 0 & \cdot & \cdot &  &  C_{2} & C_{1} & C_{2} & \cdot & \cdot &   \\
\cdot & \cdot & \cdot & \cdot &  &  & \cdot & \cdot & \cdot & \cdot &  &  \\
D_{1} & D_{2} & D_{1}  & \cdot & \cdot &  &  b_{0} & 0 & 0 & \cdot & \cdot &  \\
D_{2} & D_{1} & D_{2}  & \cdot & \cdot &  &  0 & b_{1} & 0 & \cdot & \cdot &  \\
\cdot & \cdot & \cdot & \cdot &  &  & \cdot & \cdot & \cdot & \cdot &  &    
\end{array}               
\right ),  \label{A}
\ee
%nd{displaymath}
%with the following notation:
%the 'coupling terms' are 
where $ C_{1}=-D_{1}=-2\pi n(E_{1}+E_{2})$, $ C_{2}=-D_{2}=-2\pi n(E_{1}-E_{2})$,
%whereas the terms on the principal diagonal read
whereas $a_{0}=- T t\frac{4\pi^2 n^2}{t^2}$, 
$a_{k>0}=-T\frac{t}{2}(\frac{4\pi^{2}}{t^{2}}n^{2}+\pi^{2} k^{2})$, 
and $ b_{k}=-a_{k}$, $ \forall k\geq 0 $. 
The appearance of nonzero $a_k$ terms for $k\geq 1$ is due to 
the finite spatial extension of the string.
One can prove that  
\bea
Det(A) & = & \prod_{i=1}^{\infty} a_{i}  b_{i} \times  
             \left[ 1 - C_{1}D_{1}(\sum_{(i-j)even}\frac{1}{a_{i}b_{j}})-C_{2}D_{2}
            (\sum_{(i-j)odd}\frac{1}{a_{i}b_{j}}) \right.  \nn \\ 
       &   & \left. + (C_{1}^2-C_{2}^2)(D_{1}^2-D_{2}^2) 
       ( \sum_{(i-j)odd,(k-l)odd} \frac{1}{a_{i}a_{j}b_{k}b_{l}} ) \right]. 
\eea
$\prod_{i=1}^{\infty} a_{i}  b_{i}$ corresponds to the uncoupled coordinates, 
for which we can use (\ref{free1}).
Using now the identities (valid for some complex $x$)

\bea
\frac{1}{2x^2} & + & \sum_{k=2,4,6\dots}\frac{1}{x^{2}+k^{2}}=\frac{\pi}{4x}\coth(\frac{\pi x}{2}) \label{cot} \\
               &   & \sum_{k=1,3,5\dots}\frac{1}{x^{2}+k^{2}}=\frac{\pi}{4x}\tanh(\frac{\pi x}{2}), \label{tan}
\eea
%and remembering that the path integral is given by the inverse of the determinant,
we obtain the following partition function:

\be
Z_{01}=\frac{T}{4\pi}e^{\frac{\pi}{3}\frac{1}{t}}
        \prod_{n \geq 1}
         \frac{[(1-E_1^2)(1-E_2^2)]^{-1}}
                  {[1-\frac{(1+E_1)(1+E_2)}{(1-E_1)(1-E_2)}e^{-4\pi n\frac{1}{t}}]
                   [1-\frac{(1-E_1)(1-E_2)}{(1+E_1)(1+E_2)}e^{-4\pi n\frac{1}{t}}]}.
\ee
The string tension $T$ was absorbed into $E$: $E/T\rightarrow E$.

\noindent
Using the convenient notation $\epsilon=\epsilon_1+\epsilon_2$, with
$\epsilon_j=arcthE_j, \hbox{  } j=1,2$,
as well as $\zeta$-function regularizing the divergent product 
$\prod_{n\geq 1}[(1-E_1^2)(1-E_2^2)]$ via $\sum_{k\geq 1} 1
=lim_{s\rightarrow 0}\zeta (s) = -\frac{1}{2}$, we finally obtain

\be
Z_{01}=\frac{T}{4\pi}e^{\frac{\pi}{3}\frac{1}{t}} \sqrt{(1-E_1^2)(1-E_2^2)} 
       \prod_{n=1}^{\infty}
          \frac{1}
                  {[1-e^{2\epsilon-4\pi n\frac{1}{t}}]
                   [1-e^{-2\epsilon-4\pi n\frac{1}{t}}]}. \label{z01_1}
\ee
Under the $\zeta$-function regularization, 
the divergent infinite product, to which {\it all} the string oscillation modes
along $\sigma$ do contribute, has metamorphosed into
the Born-Infeld term $\sqrt{(1-E_1^2)(1-E_2^2)}$.
This happens not only in the case of globally neutral strings \cite{bp} but rather,
as usual from T-duality/D-branes arguments,
each string end-point has associated with it a separate BI action, no matter what its charge is.

We quote now the result of a longer calculation, performed
in $D$ dimensions without previous diagonalization of $F_{\mu\nu}$. 
It says that the partition function  is just the one without
electromagnetic field present, times a product of BI-like factors:
\be
Z_{(F_{\mu\nu})}=Z_{(F=0)}\times \prod_{n=1}^{\infty}
[Det(\eta_{\mu\nu}-cth(2\pi n/t)F_{\mu\nu})]^{-1}. \label{nodiag}
\ee 
This shows the importance of the BI action for the whole problem treated here: 
(\ref{nodiag}) encodes all the effects of the external field, 
in particular the way it distorts the open string spectrum.
Eq. (\ref{nodiag}) arises from the determinant of a matrix analogous to (\ref{A})
(now containing $D \times D$ infinite blocks) upon summing over the 
$\sigma$-oscillators of the string. Remarkably enough, after extensive use of (\ref{cot},\ref{tan}),
each infinite off-diagonal block
gets replaced by a simple term $F_{\mu\nu}cth(2\pi n/t)$, whereas on the principal
diagonal one gets the metric tensor $\eta_{\mu\nu}$.

It is sometimes useful to recast (\ref{z01_1}) in a different form
(switching from the closed string channel to the open string one).
Using the transformation properties of the Dedekind eta function
$\eta(x)=e^{i\pi \frac{x}{12}}\prod_{1}^{\infty}(1-e^{2\pi inx})=
\frac{1}{\sqrt{-ix}}\eta(-\frac{1}{x})$,
%and remembering that the $\Theta_1$-function
%$\Theta_{1}(v|\tau)=2q^{\frac{1}{8}} \sin{\pi v} 
%\prod_{n=1}^{\infty}(1-q^{n})(1-e^{2\pi i v}q^{n})(1-e^{-2\pi i v}q^{n}) $
%behaves under modular transformations as follows
%satisfies the relation
%$\Theta_{1}(v|\tau)=-e^{-i\frac{v^2}{\tau}}\frac{1}{\sqrt{-i\tau}}
%\Theta_{1}(\frac{v}{\tau}|-\frac{1}{\tau})$,
and of the first $\Theta$-function
$\Theta_{1}(v|\tau)=2q^{\frac{1}{8}} \sin{\pi v} 
                    \prod_{n=1}^{\infty}(1-q^{n})(1-e^{2\pi i v}q^{n})(1-e^{-2\pi i v}q^{n})
                   =-e^{-i\frac{v^2}{\tau}}\frac{1}{\sqrt{-i\tau}}
                    \Theta_{1}(\frac{v}{\tau}|-\frac{1}{\tau})$,
(\ref{z01_1}) becomes 
\be
Z_{01}=(E_1+E_2) \frac{T}{4\pi}e^{\frac{\pi}{12}t}  
       \frac{e^{-\frac{t}{2\pi}\epsilon^2}}{sin(\epsilon \frac{t}{2})}
       \prod_{n>1}
          \frac{1}
                  {[1-e^{(i\epsilon-\pi n)t}]
                   [1-e^{-(i\epsilon+\pi n)t}]}. \label{z01_2}
\ee
A linear $(E_1+E_2)$ factor appears in front instead of the BI term. 
Both (\ref{z01_1}) and (\ref{z01_2}) display poles, which signal the string pair production. 
They are
\be
t_p(k)=\frac{2k\pi}{\epsilon}; \hbox{   } k=0,1,2,\dots \label{poles}
\ee

The path integrals along the other coupled directions are obtained from (\ref{z01_1}) or (\ref{z01_2}). 
The $Z_{23}$ path integral, for the $2$ and $3$ directions  
coupled by a magnetic eigenvalue $B$, for instance,  
is obtained  by replacing 
$\epsilon_{1,2} \rightarrow if_{1,2}$ ($i=\sqrt{-1}$ ) 
in (\ref{z01_1}) \cite{schwinger,bp}.
%\be
%\frac{1+iB_j}{1-iB_j}=e^{2if_j},
%\ee 
Now $f_{1,2}=arctgB_{1,2}$ and $f=f_1 + f_2$. $Z_{23}$ does not exhibit poles, as expected:
\bea
Z_{23}&=&\frac{T}{4\pi}e^{\frac{\pi}{3}\frac{1}{t}} \sqrt{(1+B_1^2)(1+B_2^2)} 
           \prod_{n>1}
             \frac{1}
                  {[1-e^{2if -4\pi n\frac{1}{t}}]
                   [1-e^{-2if-4\pi n\frac{1}{t}}]} \nn \\ 
                         &=&\frac{T}{4\pi}e^{\frac{\pi}{12}t} 
           (B_1+B_2) 
           \frac{e^{\frac{t}{2\pi}f^2}}{sh(f \frac{t}{2})}
             \prod_{n>1}
               \frac{1}
                  {[1-e^{ (f-\pi n)t}]
                   [1-e^{-(f+\pi n)t}]}.  \label{z_23}
\eea
%The above expressions do not exhibit poles, as expected.  

One has to take into account also the uncoupled directions and the ghosts. 
The contribution of one free coordinate, denoted $Z_{free}$,
has been displayed in (\ref{free1}, \ref{free2});  
the ghosts cancel the stringy part of two free coordinates and give
\be
z^2_{g}=(Z_{free})^{-2}\times\frac{T}{2\pi t}. \label{zghosts}
\ee
Using eqs. (\ref{free1}) or (\ref{free2}), (\ref{zghosts}), 
(\ref{z01_1}) or (\ref{z01_2}),
and (\ref{z_23}),
one can write down the whole partition function, 
$Z=z^2_g \times (Z_{free})^d \times Z_{01}\times Z_{23}\times Z_{45}\times\dots$, 
in $D$ dimensions out of which $d$ are left uncoupled by the electromagnetic field. 
%It reads:
%\bea
%Z_{(D,d)}&=&(\frac{T}{4\pi})^{D/2}e^{-(D-2)\frac{\pi}{24}\frac{t}{l}}
%            \left[ \prod_{n>1}  \frac{1}
%                         {[1-e^{(-\pi n)\frac{t}{l}}]
%                          [1-e^{-(+\pi n)\frac{t}{l}}]} \right]^{d-2} \\
%         &\times & (E_1+E_2)  e^{-\frac{1}{2\pi}\frac{t}{l}\epsilon^2} 
%         \frac{1}{sin(\epsilon \frac{t}{2l})}
%                   \prod_{n>1}  \frac{1}
%                         {[1-e^{(\epsilon-\pi n)\frac{t}{l}}]
%                          [1-e^{-(\epsilon+\pi n)\frac{t}{l}}]}  \nn  \\
%         &\times & (B_1+B_2)  e^{\frac{1}{2\pi}\frac{t}{l}f^2}  
%         \frac{1}{sh(f \frac{t}{2l})}
%                   \prod_{n>1}  \frac{1}
%                         {[1-e^{(if-\pi n)\frac{t}{l}}]
%                          [1-e^{-(if+\pi n)\frac{t}{l}}]}   \nn \\ 
%         &\times & \dots  \nn
%\eea
%The $\dots$ represent new factors due to space directions 
%coupled by further magnetic-like eigenvalues, 
%in analogy with the $(X_{2},X_{3})$ pair of coordinates.5

%\bigskip
%\noindent
\section{Pair production rate}

From (\ref{free1},\ref{z01_2},\ref{zghosts}), and evaluating $Im\int\frac{dt}{t}Z_{01}(t)$, 
one obtains the pair production rate in presence of an electric field
in D dimensions (compare to \cite{bp, ai})
\be
\gamma (E) \equiv\sum_{k=1}^{\infty}\gamma_k=\pi \sum_{k} (-)^{k+1} (E_1+E_2)e^{-k\epsilon}
\frac{T}{4\pi}\frac{\epsilon}{k\pi}[Z_{free}]^{D-2},  \label{e}
\ee
with $Z_{free}$ given by eq. (\ref{free2}), in which $t$ is replaced by 
$\frac{2k\pi}{\epsilon}$, cf. (\ref{poles}).
The term $\gamma_1$ is the dominant one.
This result gets corrected in two ways in presence of a magnetic field.
First, the electric-like eigenvalue ${\cal E}$ may change in presence
of other components of $F_{\mu\nu}$, as already discussed
(thus we assume $F_{\mu\nu}$ to be in block-diagonal form, with ${\cal E}\equiv E$).
Second, the presence of non-zero magnetic-like eigenvalues changes 
the form of the production rate, a point to which we now turn our attention. 
We consider only one non-zero magnetic eigenvalue
(the analysis proceeds identically and independently for several $B$'s).
Then each term $\gamma_k$  
in the production rate (\ref{e}) 
gets corrected,
$\gamma_k (E, B)=\gamma_k (E) \times \delta_k$,
with the following correction factor
\be
\delta_k (f) =\sqrt{1+\frac{B^2}{T^2}}\prod_{n=1}^{\infty}
              \frac{(1-x^n)^2}{(1-e^{2if}x^n)(1-e^{-2if}x^n)}
             =\frac{1}{\cos f} \prod_{n=1}^{\infty}
              \frac{(1-x^n)^2}{(1-2x^n\cos 2f +x^{2n})}. \label{mag_b}
\ee
$x$ is evaluated at the pole of order $k$: $x=e^{-4\pi/t_p(k)}$.
We now discuss the behaviour of $\delta_1 \equiv \delta$
as a function of the magnetic field.

In the absence of the BI term, $\delta \leq 1$;
the magnetic field would always decrease the PPR, 
as is the case for bosonic (Klein-Gordon) point-particles.
Nevertheless, this $\sqrt{1+(\frac{B}{T})^2}$, stringy, contribution 
triggers a qualitative change; the pair production
can be enhanced by the magnetic field,
although by a small factor: $\sqrt{2}$ at most per string end-point
if we do not allow the fields to be greater than the string tension
(if we drop this restriction the PPR increases indefinitely, as $B \rightarrow \infty$). 
Although this enhancement is quite tiny, we think  
it is somehow unexpected in a bosonic theory, and further understanding
of it would be required. 
%The fact that the change is 'fine-tuned' to be 
%of order unity might also be an interesting
%, although not very useful 'practically', feature.

\begin{picture}(300,140)(50,20)
\put(100,130){\line(1,0){100}}
\put(100,51){\line(1,0){100}}
\put(100,130){\line(0,-1){79}}
\put(200,130){\line(0,-1){79}}
\qbezier(100,130)(111,122)(113.35,51)
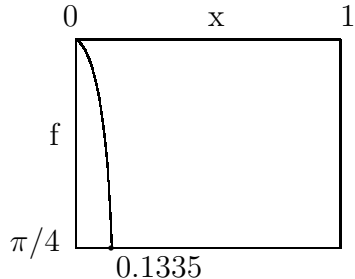
\put(70,25){fig.1  Modification of the pair creation rate due to a magnetic field}
\put(200,135){1}
\put(95, 135){0}
\put(75,50){$\pi /4$}
\put(90,90){f}
\put(150,135){x}
\put(115,40){$0.1335$}
\put(113.35,51){\circle*{2}}
\end{picture}

\noindent
In the figure above, the coordinate axes are $x=e^{-4\pi/t_p}$, $f=arctg(B)$, 
with $t_p=2\pi / \epsilon$. 
%$x=0$ corresponds to $\epsilon=\infty$, respectively $E=1$,
%whereas $x=1$ to $\epsilon=E=0$. $f=0$ means $B=0$, $f=\pi /4$
%corresponds to $B=1$. 
The part of the figure at the left of the curve connecting $(0,0)$
and $(x\simeq 0.1335, f=\pi /4)$ corresponds to an increase of the
PPR, with a maximum at $(x=0, f=\pi /4)$, where
an enhancement by a factor of $\sqrt{2}$ is obtained. 
On the above mentioned curve 
(and for $f=0$) the PPR equals the  
pure electric field one, whereas on its right it is smaller.
%For $x=1$ the production rate is zero, of course.

A dissipative, non-relativistic, limit of string theory can be obtained \cite{ij2, ai}, 
by taking the velocity $v$ of propagation of excitations along space-like coordinates
to be much smaller than the velocity of light $c=1$ along the time coordinate.
In this case, the enhancement due to the BI term may become dramatic: 
it is of the form $\sqrt{1+(\frac{B}{vT})^2}$, 
with $v \ll 1$. 
%being the above dimensionless, velocity-like, parameter.

In the case of supersymmetric strings the tachyon disappears from the free
spectrum, hence also the exponential increase it would produce (see eqs. \ref{free2}, \ref{e}).
For zero magnetic field the fermionic contribution was evaluated in \cite{bp}
and contains the sum over the even spin structures of powers of 
$\Theta$- and $\eta$-functions;
in particular, it cancels the $e^{-k \epsilon}$ factor in (\ref{e}).
An additional magnetic field further corrects the contribution of each
of the spin structures  $s$ by a factor 
\be
\delta_{k}^{(s)} (f)=
\prod_{n>0} \frac{(1+(-)^{s} 2x^n\cos 2f +x^{2n})}{(1-(-)^{s} x^n)^2}, \quad s=2,3,4, \label{mag_f}
\ee
with $n$ integer for $s=2$ and half-integer for $s=3,4$; $x$ is evaluated at the $k$-th pole, as in (\ref{mag_b}).
These corrections can increase the PPR, independently of the BI term.
This effect is just the fermionic string counterpart of what happens in the case of pointlike Dirac fermions. 
The BI factor remains unchanged and no further poles appear in the path integral.
The PPR reads
\be
\gamma_{superstring}(\epsilon,f)=\sum_{k=1}^{\infty}\gamma_{k}\delta_{k}(f)
\times \sum_{s=2,3,4} [Z_{free}^{(s)}]^{D-2} \times \delta_{k}^{(s)}(i\epsilon) \times \delta_{k}^{(s)}(f),
\ee
where $\gamma_{k}$, $\delta_{k}$, and $\delta_{k}^{(s)}$ are the ones from 
eqs.(\ref{e}), (\ref{mag_b}), and (\ref{mag_f}), respectively. $Z_{free}^{(s)}=\sqrt{\Theta_{s}(t)/\eta(t)}$
is the contribution per uncoupled fermionic direction.

We have now to remember that if we stretch the string along one direction
- in order to stabilize the theory - an additional factor 
$e^{-\frac{t_p}{2}Td^2}=e^{-\frac{\pi}{\epsilon}Td^2}$ appears.
It corresponds to a mass term $\frac{1}{2}Td^2$,
whose role is to compensate the possibly negative contribution \cite{fp, acny} of
(in string tension units):
\be
\alpha 'm^2=-\frac{f^2}{2\pi ^2}-\frac{f}{\pi}(n-1/2)+(n-1) \geq -1, \label{tachyons}
\ee
$n=a_1^{\dagger}a_1$ being the number operator for the string
modes lying on the first Regge trajectory.
Thus, it is enough to take $\frac{1}{2}Td^2=1$; this provides a damping
factor for the PPR of the form $e^{-\frac{\pi}{\epsilon}}$. Nevertheless,
for big electric fields ($E\rightarrow 1$, or $\epsilon\rightarrow \infty$)
this does not influence much the PPR, and the previous conclusions hold.

One final remark is in order. Using eqs. (\ref{z_23}) and (\ref{free2})
one sees that in $D=26$ and in the limit $t\rightarrow \infty$
the leading term in the partition function is of the form
$e^{\pi t(\frac{f^2}{2\pi^2}-\frac{f}{2\pi}+1)}$, 
modulo a prefactor. The coefficient of $(-t)$ in the exponent gives precisely
the lower tachyonic mass ($n=0$) in eq. (\ref{tachyons}), as it should.
This check confirms that the path integral encodes 
- at least in principle - 
all the information about the spectrum.
\bigskip

%We end with two speculative remarks: first, one might imagine those stretched 
%strings to lie between two brane universes, maybe playing a cosmological role
%- perhaps being a kind of dark matter. Second, the calculations presented
%here could have some relevance for QCD mesons, or for vortices in superconductors.

\bigskip
%\newpage
\noindent
{\bf Acknowledgments}

I am grateful to C. Bachas for stimulating discussions, 
in particular for suggesting the T-dual of the 'Higgs mechanism' used here.
A discussion with E. Gava is acknowledged.
I thank R. Iengo for introducing me to the subject, for very useful discussions
and encouragement, and for a careful reading of the manuscript.

%\newpage

\end{document}